\documentclass[journal=nalefd,manuscript=article]{achemso}

\usepackage{graphicx}
\usepackage{dcolumn}
\usepackage{bm}
\usepackage{epstopdf}

\title{Nanoscale electromechanics to measure thermal conductivity, expansion and interfacial losses}
\author{John P. Mathew}
\email{jomathew@tifr.res.in}
\author{Raj Patel}
\author{Abhinandan Borah}
\author{Carina B. Maliakkal}
\author{T. S. Abhilash}
\author{Mandar M. Deshmukh}
\email{deshmukh@tifr.res.in}
\affiliation{Department of Condensed Matter Physics and Materials Science, Tata Institute of Fundamental Research, Homi Bhabha Road, Mumbai 400005, India}

\begin{document}
\begin{abstract}
We study the effect of localized Joule heating on the mechanical properties of doubly clamped nanowires under tensile stress. Local heating results in systematic variation of the resonant frequency; these frequency changes result from thermal stresses that depend on temperature dependent thermal conductivity and expansion coefficient. The change in sign of the linear expansion coefficient of InAs is reflected in the resonant response of the system near a bath temperature of 20 K. Using finite element simulations to model the experimentally observed frequency shifts, we show that the thermal conductivity of a nanowire can be approximated in the 10-60 K temperature range by the empirical form $\kappa=b$T W/mK, where the value of $b$ for a nanowire was found to be $b=0.035$ W/mK$^2$, significantly lower than bulk values. Also, local heating allows us to independently vary the temperature of the nanowire relative to the clamping points pinned to the bath temperature. We suggest a loss mechanism (dissipation $\sim10^{-4}-10^{-5}$) originating from the interfacial clamping losses between the metal and the semiconductor nanostructure.

\end{abstract}

With resonant frequencies in the MHz-GHz regime and quality factors higher than $10^{6}$, nanoelectromechanical systems (NEMS) have the potential to replace traditional MEMS technologies.\cite{moser2014nanotube,chakram2014dissipation}
Resonators of thin membranes, carbon nanotubes,\cite{Sazonova2004, Huttel2009} graphene,\cite{Bunch2007} and other two dimensional materials have been studied for their use as sensors\cite{Lassagne2008, Chiu2008} and for probing fundamental physics of mechanical motion.\cite{Singh2010, Safavi-Naeini2012} In order to improve the sensitivity of the NEMS sensors microscopic mechanisms of energy loss have also been studied extensively.\cite{Mohanty2002,hutchinson2004dissipation}

Thermal properties like thermal conductivity and expansion coefficient can play an important role in the performance of NEMS devices.  Thermal stresses originating in localized heating have been little studied, and can result in variation of resonant frequencies of NEMS devices. Among many schemes of transduction, electrothermal method has been shown to be a viable option in tuning the resonant frequency of MEMS and NEMS.\cite{lammerink1990performance,syms1998electrothermal,jun2006electrothermal} Thermal stresses cause frequency shifts due to the non-zero thermal expansion coefficients and reduced thermal conductivity of nanostructures. Here, in this work, we use this sensitivity of NEMS to thermal stresses to measure the thermal conductivity of nanowire resonators and study the effect of a negative thermal expansion coefficient. Also, we measure the quality factor to study dissipation as a function of the nanowire temperature with local heating while keeping the contact metal electrodes at bath temperature. The dissipation in our resonators show strong dependence on the temperature of the contact electrodes suggesting clamping loss mechanism at the interface of the dissimilar materials.

InAs nanowires of $\sim100$ nm diameter and 10 $\mu$m length, grown using metal organic chemical vapor deposition techniques, were used in this work. We used intrinsic silicon wafers with 100 nm thermally grown nitride as the substrate. The fabrication steps are similar to the ones detailed in our previous work.\cite{solanki2010tuning} A thin layer of e-beam resist was spin coated on the substrate followed by drop casting the nanowires. Further layers of e-beam resists were spun to sandwich the nanowires. The source, drain and local gate electrodes were patterned by e-beam lithography followed by sputtering of gold (with adhesion layer of chromium) as contact electrodes and mechanical anchors. An in-situ Ar plasma was used to clean the nanowire surface prior to metal deposition. Further details on fabrication can be found in the supporting information (section A). Using intrinsic Si wafer allows us to minimize the effects of parasitic capacitances and  carry out radio frequency measurements\cite{xu2010radio} directly with a network analyzer.

\begin{figure}
\includegraphics[width=0.75\columnwidth]{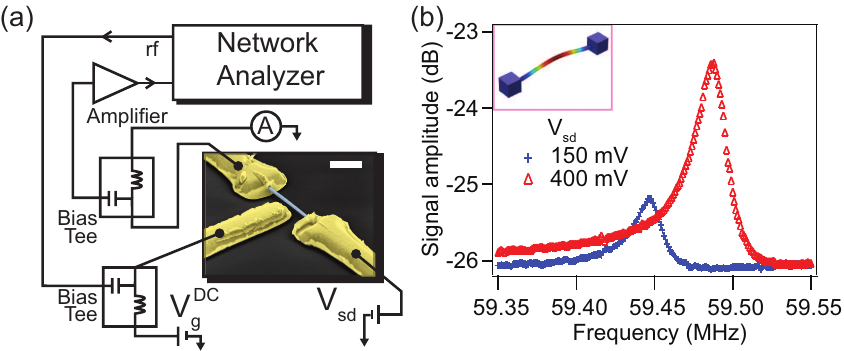}
\caption{ \label{fig:circuit} (a) Schematic diagram of the circuit used for electrical readout shown along with a false colored SEM image of the nanowire resonator device in local gate configuration. Scale bar of the image is 2 $\mu$m. (b) Resonant response of the nanowire for different source-drain bias voltages with drive power of -34 dBm at 13 K bath temperature. (Inset) Simulated mode shape of the fundamental mode of the suspended nanowire.}
\end{figure}

Figure \ref{fig:circuit}(a) shows a false colored scanning electron microscope (SEM) image of the doubly clamped nanowire resonator with local gate configuration. The nanowire diameter is 120 nm and length of the suspended region is 3.2 $\mu$m. A schematic of the measurement scheme is also shown in Figure \ref{fig:circuit}. In the capacitive scheme of actuation used here, the nanowire resonator is driven by applying a DC ($V_{g}$) and $\it{rf}$ voltage to the gate electrode. The $\it{rf}$ drive is kept sufficiently low to ensure linear response of the oscillator. As the $\it{rf}$ frequency is swept, the  power transmitted ($S_{21}$) through the gate and nanowire system shows change at resonance and is detected using a network analyzer. The signal coming from the nanowire is split into its DC and $\it{rf}$ components using a bias tee and the DC current through the nanowire is directly measured using a current amplifier. Typical resonance curves of the device are shown in Figure \ref{fig:circuit}(b). At low temperatures the quality factor of the resonator is $2-3\times10^{3}$ allowing us to measure small frequency shifts accurately. We investigate the effects of Joule heating on the fundamental mode frequency of the resonator at cryogenic temperatures. Figure \ref{fig:circuit}(b) inset shows the fundamental mode shape obtained from simulation with finite element method (FEM).

\begin{figure}[]
\includegraphics[width=0.75\columnwidth]{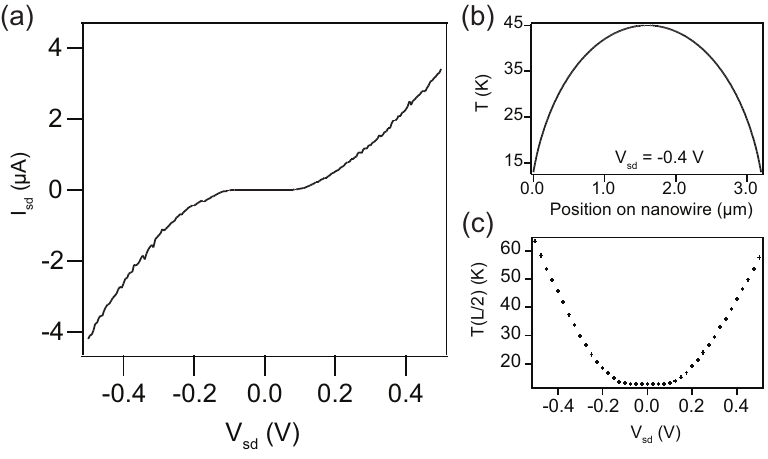}
\caption{ \label{fig:temp rise} (a) Measured I-V response of the nanowire at 13 K bath temperature with V$_{g}=-20$ V. (b) Temperature profile along the length of the nanowire obtained from simulation for a bias of V$_{sd}=-0.4$ V. (c) Maximum temperature in the nanowire shown as a function of the applied bias voltage obtained from FEM simulations done using COMSOL. The temperature rise in the nanowire can be many times the bath temperature due to low thermal conductivity of the nanowire. Comparison of simulation and analytic results is shown in section B of the supporting information. }
\end{figure}

Figure \ref{fig:temp rise}(a) shows the current-voltage response of the nanowire at a fixed gate voltage. The presence of a DC current leads to Joule heating in the nanowire. Due to the low thermal conductivity of these nanostructures\cite{Li2003a,Hochbaum2008,Boukai2008,Dhara2011b} considerable amount of heat accumulates in the nanowire even though the clamps are pinned to the bath temperature using gold electrodes; consequently the temperature of the nanowire increases while developing a spatial profile (discussed next). The heat flow along the length ($L$) of a nanowire of cross sectional area $A$ and thermal conductivity $\kappa$ is governed by the one dimensional heat equation given by:
\begin{equation}
A\frac{d}{dx}(\kappa\frac{dT}{dx})+\dot{w}=0
\label{heat}
\end{equation}
where $\dot{w}=I^{2}R/L$ is the rate of Joule heating per unit length for a current $I$ passing through the nanowire of resistance $R$. Here $\kappa$ is temperature dependent and assumed to be of the form $\kappa=b$T where $b$ is a constant. As individual nanowires can have significant variations in thermal conductivity we assume this empirical form of $\kappa$ for the temperature range of 10-60 K. This empirical form is supported by previous measurements of thermal conductivity of silicon and InAs nanowires that show near linear temperature dependence of thermal conductivity at cryogenic temperatures.\cite{hochbaum2008enhanced,zhou2011thermal,Dhara2011b}  Equation \ref{heat} is solved to obtain the temperature profile along the nanowire for an applied $V_{sd}$ assuming the contacts are in thermal equilibrium with the bath temperature ($T(x)=T_{b}$ for $x=0,L$). The temperature profile along the nanowire can be analytically written as:
\begin{equation}
T(x)=\sqrt\frac{A b T_{b}^2+L\dot{w}x-\dot{w} x^2 }{A b}
\label{T_profile}
\end{equation} where $x$ takes values between 0 and L. The assumption of thermal equilibrium with the bath temperature at the contacts is supported by the following observations: (i) the gold electrodes are much larger than the nanowire and a considerable portion ($>$ 1 $\mu$m) of the nanowire is mechanically held inside the gold contacts, and (ii) the electrical and thermal conductivity of the metal electrodes are orders of magnitude larger than that of the nanowire. For typical DC currents of few micro-Amperes sent through the nanowire, temperature rise at the contacts is negligible.

We also performed FEM simulations using COMSOL to investigate the effect of Joule heating on the temperature and mechanical properties of the nanowire. For $b=0.035$ W/mK$^2$ we see (Figure \ref{fig:temp rise}(c)) a temperature of 60 K in the middle of the nanowire for a bias voltage of 500 mV. This matches well with the analytic form of $T(x)$ as given in equation \ref{T_profile}. The procedure for obtaining $b$ by simulating the experimentally observed frequency shift with heating is described later in the manuscript. As the value of $\kappa$ is low, few micro-Amperes of current can drive the temperature in the nanowire to many times the base temperature and change the stress state of the resonator. This presents an inherent limit to carrying out electrical measurements on suspended devices of materials with low thermal conductivity.

The nanowire is initially under a high tensile stress of $\sim0.3$ GPa at low temperatures. This stress arises from fabrication processes and thermal expansion of the substrate, electrodes, and nanowire while cooling to low temperatures. We studied the change in this stress state, as reflected in the resonant frequency, with Joule heating. As the nanowire is Joule heated, a non-zero thermal expansion coefficient causes an additional stress in the nanowire leading to a change in tension in these doubly clamped structures. Here the substrate and electrodes remain at the bath temperature and, hence, do not contribute with any relative expansion to the nanowire. As the resonant frequency of the nanowire is strongly dependent on tension, any change will cause a shift in the resonant frequency. Previous reports have shown a decrease in resonant frequency with Joule heating in systems with positive thermal expansion coefficient. \cite{remtema2001active, karabalin2009parametric,Abhilash2012} At room temperature we see a monotonic decrease in resonance frequency of the InAs nanowire resonator when heated (supporting information section A) due to its positive expansion coefficient.

\begin{figure}
\includegraphics[width=0.75\columnwidth]{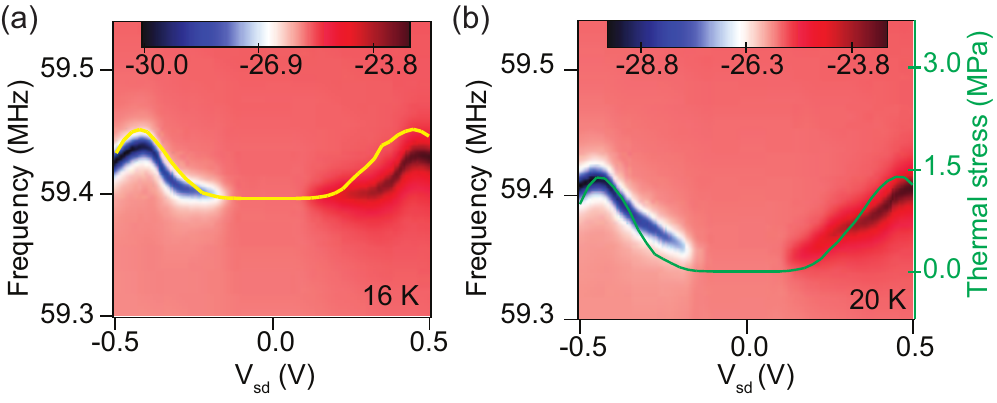}
\caption{ \label{fig:dispersion} (a) Experimentally observed variation of resonant frequency of the nanowire with source-drain bias voltage at 16 K bath temperature. The color scale units are in dB. The DC current through the nanowire increases with increasing bias voltage and contributes to Joule heating in the nanostructure that leads to a thermal stress. The nature of the frequency shift is related to the sign of thermal expansion coefficient of InAs. For a positive expansion coefficient the tensile stress reduces and the resonant frequency decreases. The yellow line is the frequency shift obtained from simulations. (b) Frequency variation with V$_{sd}$ obtained at 20 K bath temperature. The color scale units are in dB. The green line shows the additional stress (right axis) induced in the nanowire due to Joule heating obtained from simulations.}
\end{figure}

Figure \ref{fig:dispersion} shows the change in resonant frequency of the nanowire with $V_{sd}$ at cryogenic bath temperatures. At 16 K bath temperature we observe a shift in the nanowire resonance showing both positive and negative change. This is due to the non-monotonic dependence of linear thermal expansion coefficient of InAs on temperature. Certain materials, including InAs, are known to have a region of negative thermal expansion at low temperatures (10-50 K).\cite{Sparks1967} As the temperature in the nanowire varies from the bath temperature to about 60 K due to Joule heating, regions of expansion and contraction are set up in the nanowire which sets up a net thermal stress on the nanowire. The local nature of Joule heating allows us to study the effects of thermal stress solely on the nanowire, whereas in global (bath) heating, the effects of expansion of the electrodes and substrate also have to be accounted for (supporting information section A).

We extended our simulations to calculate the amount of thermal stress in the nanowire due to thermal expansion from Joule heating. We modeled the nanowire as a cylindrical beam with rigid endpoints under an initial tensile stress. The initial stress was used as a simulation parameter to match the resonant frequency at zero bias voltage. (More details about the simulations are available in section B of the supporting information.)

The DC current measured from experiments and the expansion coefficient of InAs was used as input for the simulations and the thermal conductivity of the structure was chosen to be of the form $\kappa=b$T. Figure \ref{fig:dispersion} (a) shows the simulated change in resonant frequency of the nanowire with applied $V_{sd}$. We carried out simulations for different values of $b$ and the simulated frequency shift with bias voltage was compared with the experimentally observed shift at various bath temperatures. The value of $b$ that minimized the difference for this device at all bath temperatures was obtained to be $b=0.035\pm0.010$ W/mK$^2$. We see that the value of thermal conductivity at cryogenic temperatures is close to previously reported value of $\kappa$ for InAs nanowires.\cite{zhou2011thermal,Dhara2011b}

To get a physical understanding we also develop an analytic model that describes our observations. For a tensile stress $\tau$ on a resonator of length $L$, cross sectional area $A$, and inertia moment $I$, the natural frequency is given by\cite{tilmans1992micro}:
\begin{equation}
f_{0}=\left(\frac{1}{2\pi}\sqrt{\frac{E I \beta^4}{\rho A}}\right)\sqrt{1+\frac{0.55 \tau A}{\beta^2 E I}}
\label{resonance}
\end{equation}
where $\beta=4.73/L$ is a mode factor for the fundamental mode, $E$ is the Young's modulus (97 GPa) and $\rho$ is the mass density of InAs. The stress can be written as $\tau=\tau_{0}+\Delta\tau_{h}$ where $\tau_{0}$ can arise from post fabrication stresses on the nanowire and variation of thermal strain as the device is cooled. The additional thermal stress $\Delta\tau_{h}$ accumulates on the nanowire due to relative expansion and contraction when Joule heated. As the thermal expansion coefficient is strongly temperature dependent we obtain the longitudinal thermal stress by integrating the stress over the length of the nanowire using:
\begin{equation}
\Delta\tau_{h}=-\frac{E}{L}\int_0^L\left[\int_{T_{b}}^{T(x)}\alpha[T(x)] dT \right]dx
\label{stress}
\end{equation}
where $T_{b}$ is the bath temperature and $\alpha$ is the temperature dependent expansion coefficient obtained from literature.\cite{Sparks1967} Figure \ref{fig:dispersion}(b) shows the variation of resonant frequency of the nanowire resonator at bath temperature of 20 K. The green line shows the simulated thermal stress. Simulations show that the additional stress that is induced on the nanowire is of the order of 1 MPa. Data from another nanowire device is given in section C of the supporting information.

Thus far we have discussed how controlled heating and measurement of resulting thermal stresses can allow measurement of thermal conductivity and reflect the non-monotonicity of thermal expansion of InAs. Our device geometry also allows us to address the question of how damping losses of the system change if the local temperature of the clamps is varied while the average temperature of the suspended nanowire is maintained to be constant. We find that there is a loss mechanism that depends on the temperature of the dissimilar material (InAs and gold) interface.

\begin{figure}[h]
\includegraphics[width=0.4\columnwidth]{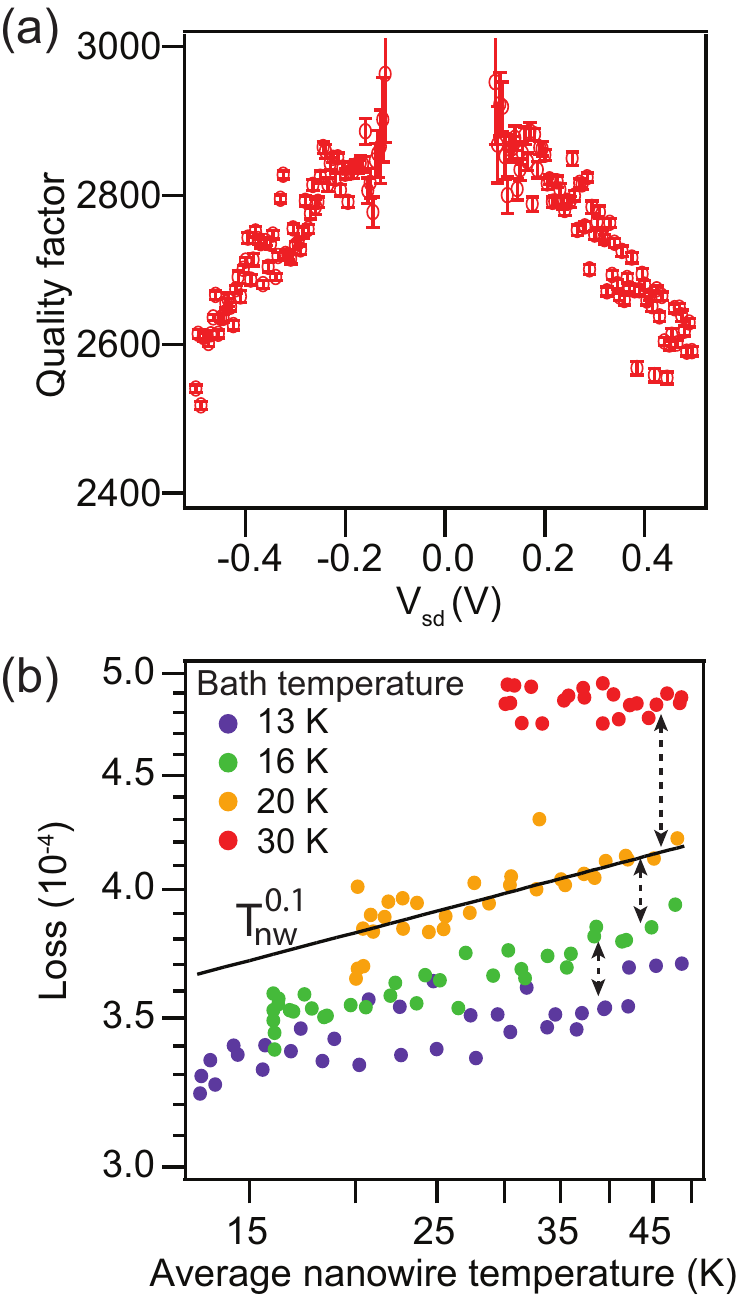}
\caption{\label{fig:Qfactor} (a) Measured quality factor of the resonator as a function of applied bias voltage at 16 K bath temperature. The quality factor reduces due to increase of the average nanowire temperature with Joule heating. (b) Inverse of the resonator quality factor (loss) is plotted as a function of the average nanowire temperature at various bath temperatures (log-log plot). The trend is seen to be same at lower bath temperatures with the exception of a positive offset (dotted lines) with increasing bath temperature.  The average nanowire temperature is obtained from simulations.}
\end{figure}

The energy loss in the resonator can be quantified by the quality factor Q. This energy loss can be due to various factors such as thermoelastic dissipation, surface losses, viscous damping, and clamping losses.\cite{lifshitz2000thermoelastic,Mohanty2002,unterreithmeier2010damping} The quality factor was extracted and studied as a function of the applied bias voltage and the average temperature of the nanowire. The Q values of the nanowire are plotted as a function of bias voltage at 16 K in Figure \ref{fig:Qfactor}(a). We see that Q decreases with $|V_{sd}|$ due to Joule heating. To further elucidate the role of Joule heating on the dissipation in the system we study the experimentally observed loss $(1/Q)$ as a function of the average temperature of the nanowire obtained from simulations.

Figure \ref{fig:Qfactor}(b) shows the variation of loss in the nanowire as a function of the average temperature of the nanowire at different bath temperatures. We see that the loss increases with increasing average temperature of the nanowire (T$_{nw}$). The variation of $1/Q$ with T$_{nw}$ is seen to follow a power law dependence (T$_{nw}^{0.1}$) at lower bath temperatures. Similar power law temperature dependence has been observed in systems of GaAs, Si, and diamond.\cite{Mohanty2002,hutchinson2004dissipation} The origin of this complex temperature dependence is not fully understood. In addition, we observe that the loss at 13 K bath temperature is always lower than the loss at 30 K even when the average temperature of the nanowire is the same.

This gives insight into a potential loss mechanism in the nanowire resonator. From the offset in loss seen in Figure \ref{fig:Qfactor}(b) it is clear that the dissipation in our resonator is decided by the bath temperature. We see that this additional mechanism gives rise to a (bath) temperature dependent dissipation less than $10^{-4}$. Since the metal supports of the suspended structure remain at the bath temperature this is an indication of mechanical losses occurring at the metal clamps. Clamping losses arise when part of the vibrational energy is dissipated as a wave transmitted into the supports and the substrate. This has been studied in detail for single crystal NEMS structures where the supports and the resonator are fabricated from the same material.\cite{cross2001elastic} These losses arise purely from the geometry of the system and is temperature independent. Since in our device the resonator material is different from the clamping metal, we believe an additional loss mechanism contributes due to the gold contacts. When the nanowire oscillates, the energy lost to the clamps couples to the gold electrodes. The temperature dependent loss of gold is then reflected in the measured quality factor of the resonator.

The observation of significant loss at the clamps is further supported by the fact that our nanowire is under high tensile stress. Clamping losses are known to be the dominant damping mechanism in doubly clamped beams under high pre-stress.\cite{schmid2008damping} Here the tension should be more than the flexural rigidity such that the beam mimics the mechanics of a stretched string. This condition is satisfied when $\tau_0\gg E I /(A L^2)$. The pre-stress in our nanowire at 30 K is 313 MPa, significantly larger than $\frac{E I}{A L^2}=9$ MPa. This indicates that clamping losses in our resonator are significant at low temperatures and the temperature dependent offset observed could result from the losses at the gold clamps. Internal friction in gold films of thickness comparable to the electrodes deposited for our resonator have been previously shown to have temperature dependent losses in the range of $10^{-4}$ at low temperatures.\cite{liu1999low} This matches well with the observed offset seen in our gold clamped device. This also suggests the possibility of improving quality factor in similar NEMS structures by using low loss metals as clamps.

Other intrinsic temperature dependent loss mechanisms of the nanowire like internal defects are ruled out by the offset observed in Figure \ref{fig:Qfactor}(b) despite the average temperature of the nanowire being same.\cite{Mohanty2002} The role of viscous damping in the observed offset at different temperatures is also ruled out as the sample is cryopumped to pressures lower than 0.01 Pa.\cite{schmid2008damping} The quality factor that we obtain in our system also rules out the role of thermoelastic dissipation.\cite{schmid2008damping} The dissipation through thermoelastic damping in systems of similar dimensions is known to be orders of magnitude lower than the dissipation we observe of 4$\times$$10^{-4}$.\cite{lifshitz2000thermoelastic} One factor that could offset the quality factor with bath temperature is the change in residual stress itself. It is known that stress can be used to change the quality factor in NEMS systems.\cite{Verbridge2006,schmid2008damping} However, the change in $\tau_0$ from 13 K to 30 K bath temperature is less than 2$\%$, whereas the Q factor changes by 30$\%$ (see Figure \ref{fig:Qfactor}(b)).


In summary, we have directly probed the change in the sign of thermal expansion coefficient of a material by studying the non-monotonic variation of resonant frequency of an InAs nanowire resonator with local heating. This also suggests that the effect of Joule heating on NEMS can be negated by appropriately layering the device with materials of positive and negative thermal expansion. We obtain the thermal conductivity of the nanowire to be 0.035T W/mK$^2$ (where T is in units of K and in the range 10-60 K) demonstrating the usefulness of this technique in measurement of thermal conductivity of individual semiconducting nanostructures. We have also shown that the dissipation in the resonator increases with increasing temperature of the nanowire. Moreover, the role of an  additional dissipation channel was shown in our resonator which could be due to interfacial clamping losses. Such a loss mechanism could play a dominant part in similar metal clamped NEMS at low temperatures.

\subsection*{Acknowledgment}
We acknowledge the contribution of Mr. Mahesh Gokhale and Prof. Arnab Bhattacharya, TIFR, for growth of the nanowires. We acknowledge funding from the Department of Atomic Energy, and Department of Science and Technology of the Government of India through the Swarnajayanti Fellowship. Additional funding support from AOARD (Grant No. 124045) is also acknowledged.

\begin{suppinfo}
Data from second device, comparison between simulation and analytic solution,  additional data and figures.
\end{suppinfo}

\bibliography{InAs_expansion_bib_short}

\end{document}